\documentclass{elsart}
\usepackage{graphicx}
\journal{Astroparticle Physics}
\begin{document}

\begin{frontmatter}

\title{Comparison of simulated longitudinal profiles of hadronic air showers with MASS2 balloon data.}

\author[lpnhe]{J. Guy},
\author[lpnhe]{P. Vincent}
\author[lpnhe]{J.P. Tavernet}
\author[lpnhe]{M. Rivoal}
\address[lpnhe]{Laboratoire de Physique Nucl\'eaire et des
Hautes Energies,\\  IN2P3-CNRS et Universit\'es de Paris 6-7.}

\begin{abstract}
%%%%%%%%%%%%%%%%%%%%%%%%%%%%%%%%%%%%%%%%%%%%%%%%%%%%%%%%%%%%%%%%%%%%%%%

The KASKADE and CORSIKA air shower generators are compared to the data 
collected by MASS2 balloon experiment in 1991. The test of longitudinal profile 
for proton, helium and muon flux production provide good constraints on these 
air shower generators. KASKADE and CORSIKA especially with the new simulator 
UrQMD for low energies are found to fit these data well. This study is limited 
to a comparison of longitudinal profiles and therefore does not provide constraints 
on the overall shower development.

%%%%%%%%%%%%%%%%%%%%%%%%%%%%%%%%%%%%%%%%%%%%%%%%%%%%%%%%%%%%%%%%%%%%%%%
\vspace{1pc}
\end{abstract}
\end{frontmatter}

\section{Introduction}

%%%%%%%%%%%%%%%%%%%%%%%%%%%%%%%%%%%%%%%%%%%%%%%%%%%%%%%%%%%%%%%%%%%%%%%%%%%%%%%%%%%%%%%%%%%%%%%%%%%%%%%%%%%%
High energy gamma rays arriving at the earth interact early in the atmosphere to produce large electromagnetic air 
showers. The understanding of the shower development is of primary importance for the definition of future
high energy cosmic ray experiments, as for example the HESS Cherenkov array, currently under construction 
in Namibia \cite{hess}. In such an experiment, the Cherenkov light from the showers creates an 
image in the focal plane of a telescope. It allows an estimation of the energy, direction and impact parameter for
gamma ray induced air showers, and permits cosmic ray rejection. 

This technique is strongly correlated to our ability to predict, for incoming gamma rays of different energies, 
the amount of 
Cherenkov light generated as a function of altitude and their propagation to the observation level. 
Moreover, several sources of background have been identified. The main one comes from the hadronic component of the 
cosmic rays. A hadron arriving in the atmosphere creates an hadronic shower in which muons and neutral 
pions are produced. 
Due to their initial energy, the muons radiate Cherenkov light along their full path and their life times are long 
enough for them to hit the ground. The neutral pions decay essentially into two photons and create an 
electromagnetic component in the hadron shower.

Even if a large part of these backgrounds can be identified using off-line analysis methods 
and if the muon signal does not survive the multi-telescope coincidence, 
the consequence of this background is an increased trigger rate for the experiment, inducing 
deadtime in the data acquisition. 
For this reason, the knowledge of the hadronic background is of primary importance.
For this purpose, atmospheric shower generators have been developed over many years. The HESS collaboration currently uses 
KASCADE \cite{kascade}, CORSIKA \cite{corsika}, MOCCA \cite{mocca} and the ALTAI program \cite{altai} developed at the Max-Planck Institute of Heidelberg. 
In this paper we focus on the two first generators and we take advantage of the data from MASS2 \cite{mass2muons,mass2protons}
 balloon to test these hadronic shower generators. 
The comparison is made on the muon, proton and helium fluxes measured by MASS2 as a function of the
altitude. Even if we do not check all processes in the hadronic shower development, the muon flux is directly 
correlated to the charged meson production and decay, mainly into pions and kaons, in the various processes 
involved in the hadronic interactions. 
The few measured values of proton and helium fluxes also constrain the interaction process of the primary 
particles and the production rate of secondary protons
and helium nuclei generated by the heavy ion components of cosmic rays.

In the first three sections we describe the data taken by the MASS2 collaboration, the setup of the different
generators tested in this paper and the analysis method. The next section is dedicated to  the comparison 
and the systematical uncertainties. We then discuss these results and give our conclusions.
%%%%%%%%%%%%%%%%%%%%%%%%%%%%%%%%%%%%%%%%%%%%%%%%%%%%%%%%%%%%%%%%%%%%%%%%%%%%%%%%%%%%%%%%%%%%%%%%%%%%%%%%%%%%

\section{MASS2 experiment}

%%%%%%%%%%%%%%%%%%%%%%%%%%%%%%%%%%%%%%%%%%%%%%%%%%%%%%%%%%%%%%%%%%%%%%%%%%%%%%%%%%%%%%%%%%%%%%%%%%%%%%%%%%%%
The MASS2 balloon experiment was launched at Fort Sumner, New Mexico in September 1991. 
It was composed of a superconducting magnet spectrometer, a time of flight chamber, a Cherenkov detector and 
an imaging calorimeter. 
These devices permitted cosmic-ray muon, proton and helium identification and energy reconstruction 
for several atmospheric depths from 5 g.cm$^{-2}$ to 886 g.cm$^{-2}$ \cite{mass2muons,mass2protons}. 

In this analysis we have used the following measurements: the momentum spectra of $\mu^-$ within  
different atmospheric depth ranges,  $\mu^-$ and  $\mu^+$ fluxes as a function of 
atmospheric depth respectively in the 0.3-40 GeV/$c$ and 0.3-1.5 GeV/$c$ momentum ranges, the proton 
and helium spectra at the top of atmosphere (5 g.cm$^{-2}$) and their flux variation with atmospheric 
depth in the 4.1-14.1 GeV/$c$ and 1.7-9.1 GeV/nucleon kinetic energy range respectively.

These data permit hadronic air-shower models to be tested. From proton and helium energy spectra measured at 
the top of atmosphere, we can simulate their hadronic interaction and air-shower development to compare the 
estimated muon, proton and helium fluxes with those measured at lower altitudes.
 
This study depends on the knowledge of geomagnetic field and a parametrization of atmosphere.
The former was taken from the National Geophysical Data Center \cite{NGDC},  
we have used  B$_{\mathrm{x}}$ = 23.7 $\mu$T and B$_{\mathrm{z}}$ = 45.5 $\mu$T for the Fort Sumner site. 
Concerning the latter, no specific measurement from this site was available. 
Several radiosonde observations were performed \cite{radiosonde} in the vicinity, for example 
at Amarillo and Midland-Odessa (Texas) and Albequerque (New Mexico).
All results in this paper were obtained with the parametrization from Midland-Odessa data.
The resulting uncertainties are discussed in section 6. 

%%%%%%%%%%%%%%%%%%%%%%%%%%%%%%%%%%%%%%%%%%%%%%%%%%%%%%%%%%%%%%%%%%%%%%%%%%%%%%%%%%%%%%%%%%%%%%%%%%%%%%%%%%%%

\section{Analysis}

%%%%%%%%%%%%%%%%%%%%%%%%%%%%%%%%%%%%%%%%%%%%%%%%%%%%%%%%%%%%%%%%%%%%%%%%%%%%%%%%%%%%%%%%%%%%%%%%%%%%%%%%%%%%
Proton, helium, and heavier nuclei cosmic rays were injected at 5 g/cm$^2$ in the air-shower simulation programs.
Proton and helium spectra were fitted from MASS2 data in the 4-100 GeV/c momentum range and 1.6-50 GeV/nucleon respectively.
We took into account the low energy curvature and assumed a power-law limit at higher energies with a slope of $-2.7$ 
for protons and $-2.65$ for helium (\cite{ICRCslopes}) up to 50 TeV/c.
The heavier nuclei were generated using abundances from the Particle Data Group \cite{cosmicPDG} and power-law spectra with 
slope -2.70. A sample of 10,000 showers per hadronic model were simulated.

Since muons suffer sizable multi-scattering, we generated primary particles isotropically 
from $0^{\circ}$ up to $60^{\circ}$ zenith angle, 
but only those secondary muons with angles up to $30 ^{\circ}$ were used for the flux calculations. 
The fluxes at zenith were then derived from the integrated 
flux in this interval assuming a $\cos^{2.02}(\theta)$ angular dependance for muons \cite{wentz_saltlake}.    

In the following, all fluxes were normalized to the proton and helium fluxes at 5 g/cm$^2$.
 
%%%%%%%%%%%%%%%%%%%%%%%%%%%%%%%%%%%%%%%%%%%%%%%%%%%%%%%%%%%%%%%%%%%%%%%%%%%%%%%%%%%%%%%%%%%%%%%%%%%%%%%%%%%%

\section{Generators setup}

%%%%%%%%%%%%%%%%%%%%%%%%%%%%%%%%%%%%%%%%%%%%%%%%%%%%%%%%%%%%%%%%%%%%%%%%%%%%%%%%%%%%%%%%%%%%%%%%%%%%%%%%%%%%

KASCADE \cite{kascade} hadronic interactions are based on a model from Gaisser and Stanev. 
The momentum of a leading particle is computed along with the multiplicity and momenta of resulting pions. 
The model parameters are based on hadron-nucleus and $\pi-$nucleus collision data.
This generator is used by atmospheric Cerenkov detectors such as CAT \cite{cat} and CELESTE \cite{celeste}.
CORSIKA is extensively used in high energy cosmic ray experiments and also by Cerenkov detectors as HEGRA \cite{hegra}.
CORSIKA allows several hadronic models. In version 5.60, ISOBAR and GHEISHA \cite{gheisha}, 
were proposed for low-energy interactions 
(E$_{lab}<$80 GeV for GHEISHA and 50 GeV for ISOBAR) and five other models for higher energies 
VENUS \cite{venus}, QGSJET \cite{qgsjet}, 
DPMJET \cite{dpmjet}, SIBYLL \cite{sibyll} and HDPM \cite{hdpm}. CORSIKA has recently been upgraded 
to provide modified versions of hadronic packages 
and new ones, UrQMD \cite{urqmd} and NEXUS \cite{nexus} respectively for low and high energy interactions.

%%%%%%%%%%%%%%%%%%%%%%%%%%%%%%%%%%%%%%%%%%%%%%%%%%%%%%%%%%%%%%%%%%%%%%%%%%%%%%%%%%%%%%%%%%%%%%%%%%%%%%%%%%%%

\section{Results}

\begin{figure}
\begin{minipage}[]{7cm}
\includegraphics[width=7cm]{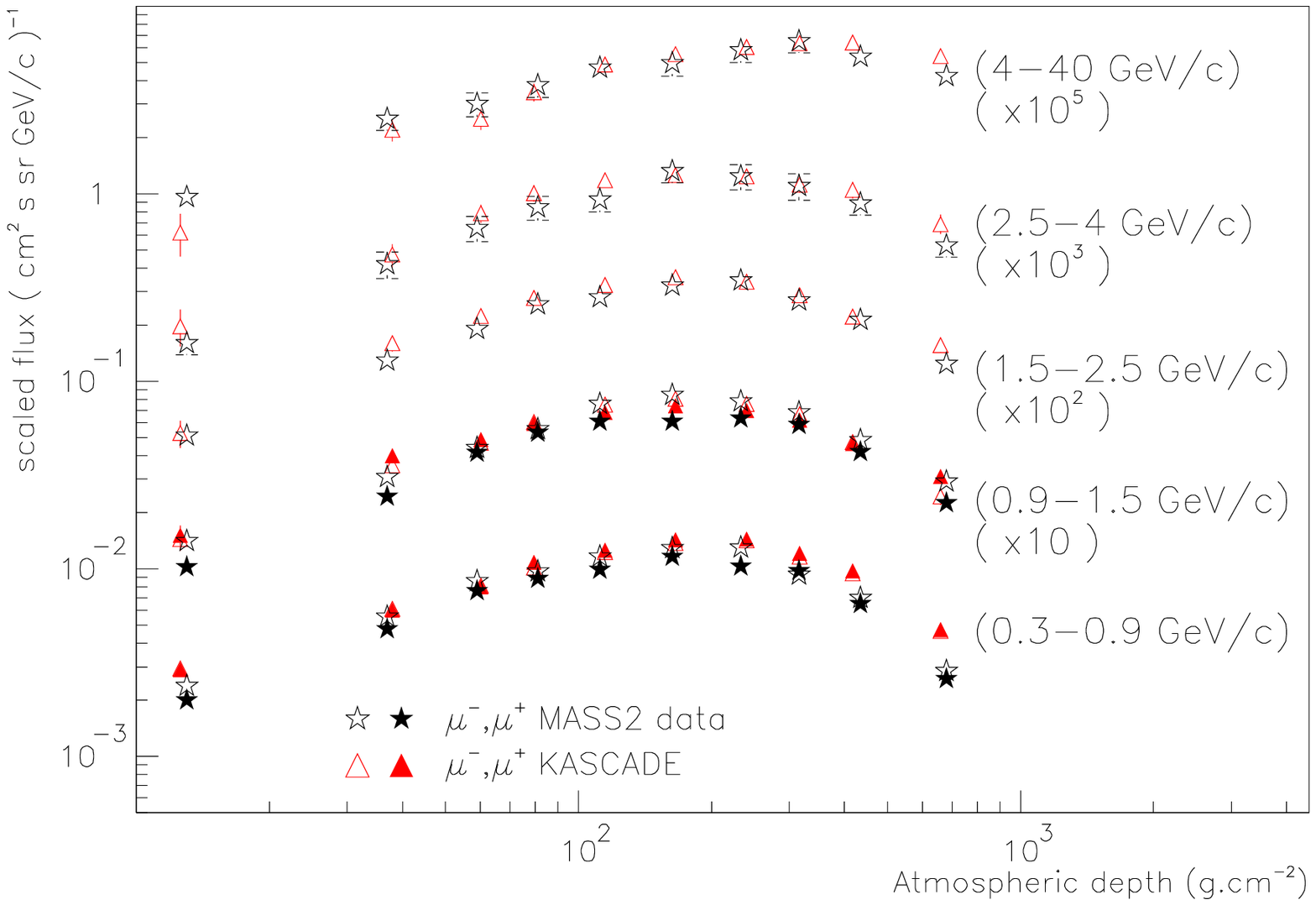}
\end{minipage}
\begin{minipage}[]{7cm}  
\includegraphics[width=7cm]{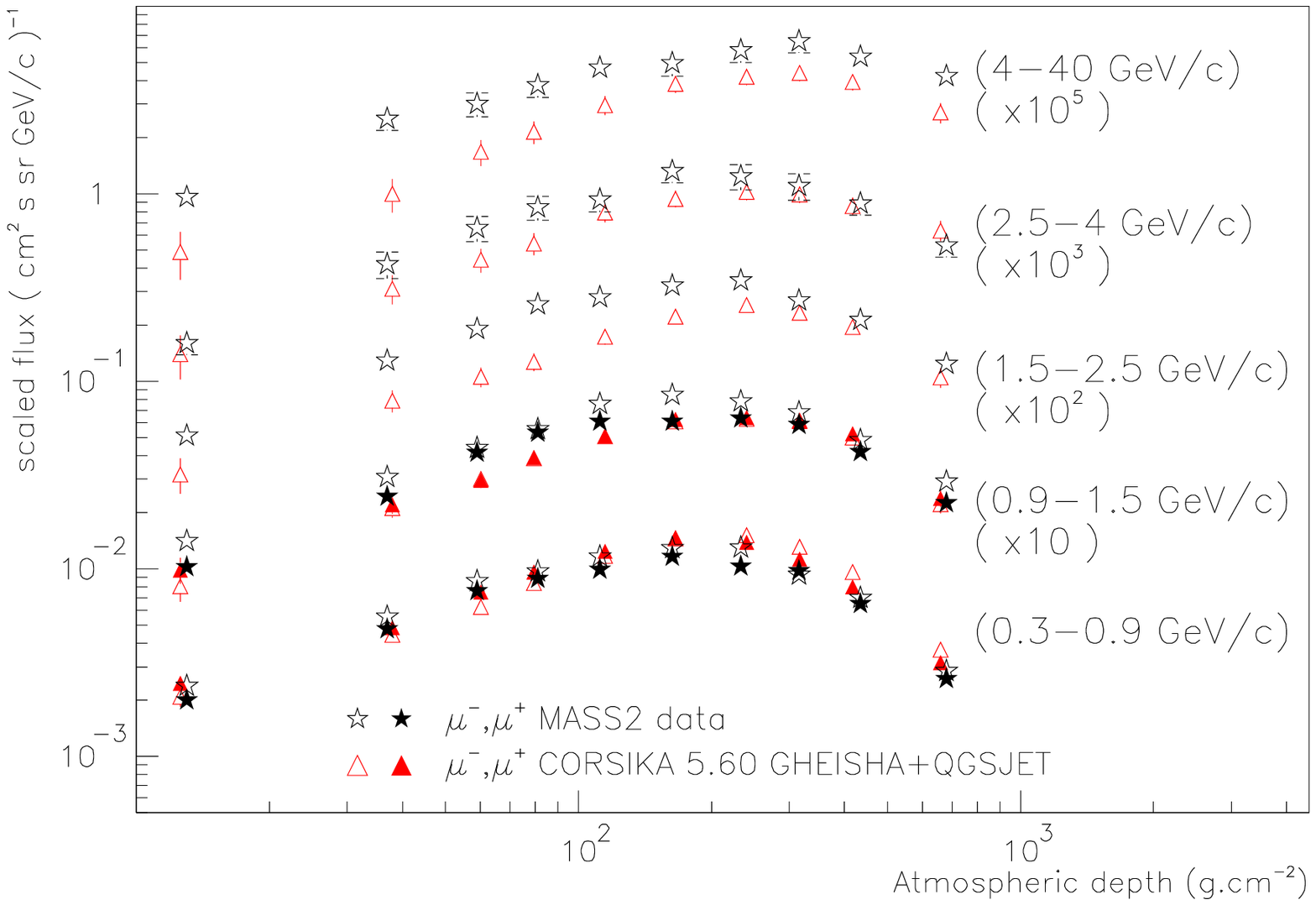}
\end{minipage}
\begin{minipage}[]{7cm}
\includegraphics[width=7cm]{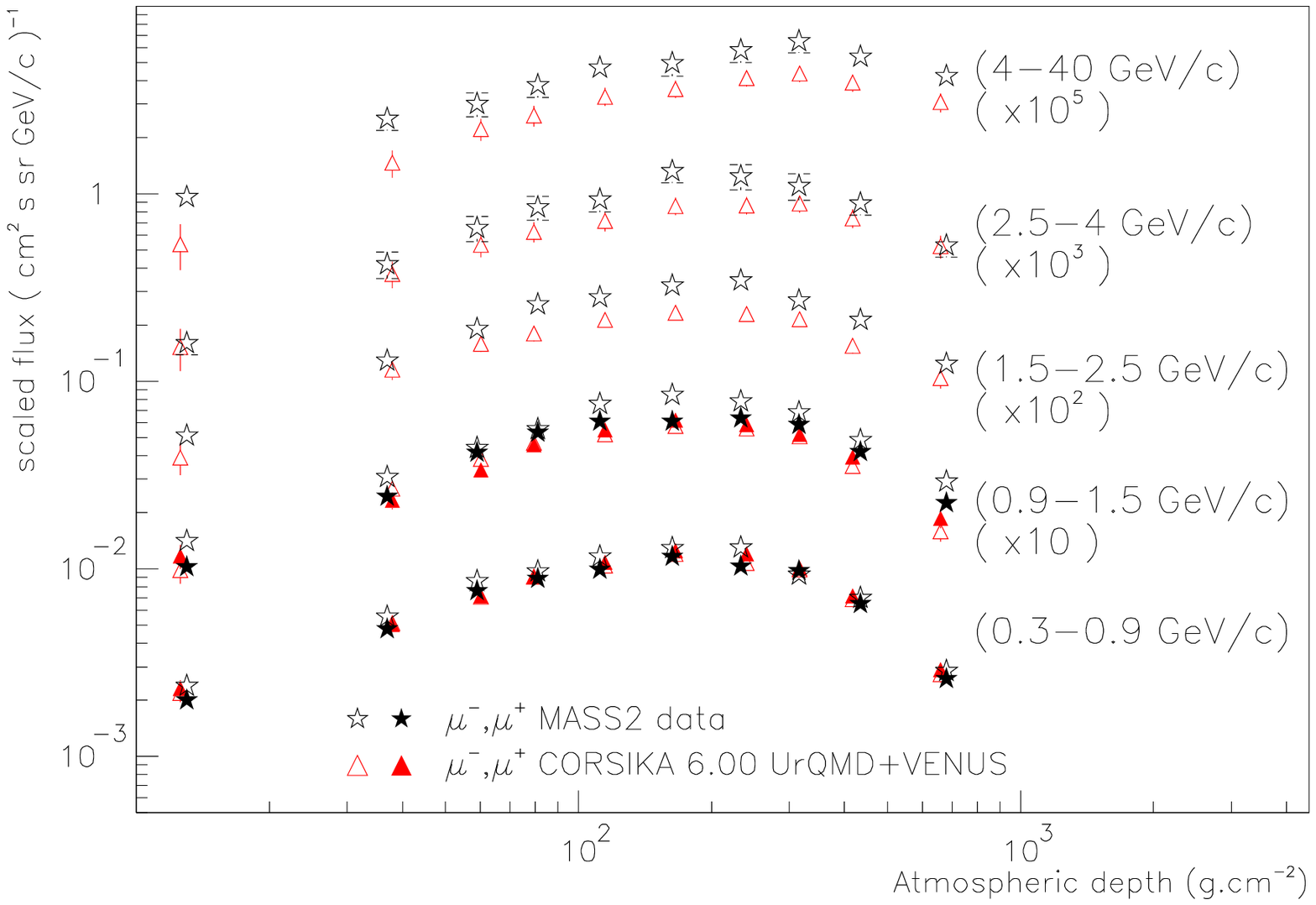}
\end{minipage}
\caption{
$\mu^{-}$ and $\mu^{+}$ fluxes as a function of atmospheric depth for different 
momentum intervals measured by MASS2 \cite{mass2muons} (stars) compared with simulations. 
Some of the distributions have been scaled.
} \label{muondepth}
\end{figure}

All available hadronic models are tested. We compute a $\chi^2$ which is the quadratic 
sum of the differences between the MASS2 \cite{mass2muons,mass2protons}
fluxes and the predictions of the simulations divided by an error term which is the 
quadratic sum of MASS2 error and the simulation statistical uncertainty. 
The MASS2 data points used in this $\chi^2$ are the fluxes measured at several 
atmsopheric depths for a given energy range. Those points are shown 
figure \ref{primarydepth} for proton and helium and figure \ref{muondepth} for muons. 
In the latter, five energy ranges are considered.
The $\chi^2$ values obtained for all models are listed table  \ref{table_chi2}.

\subsection{Protons and Heliums}

\begin{figure}
\includegraphics[width=14.cm]{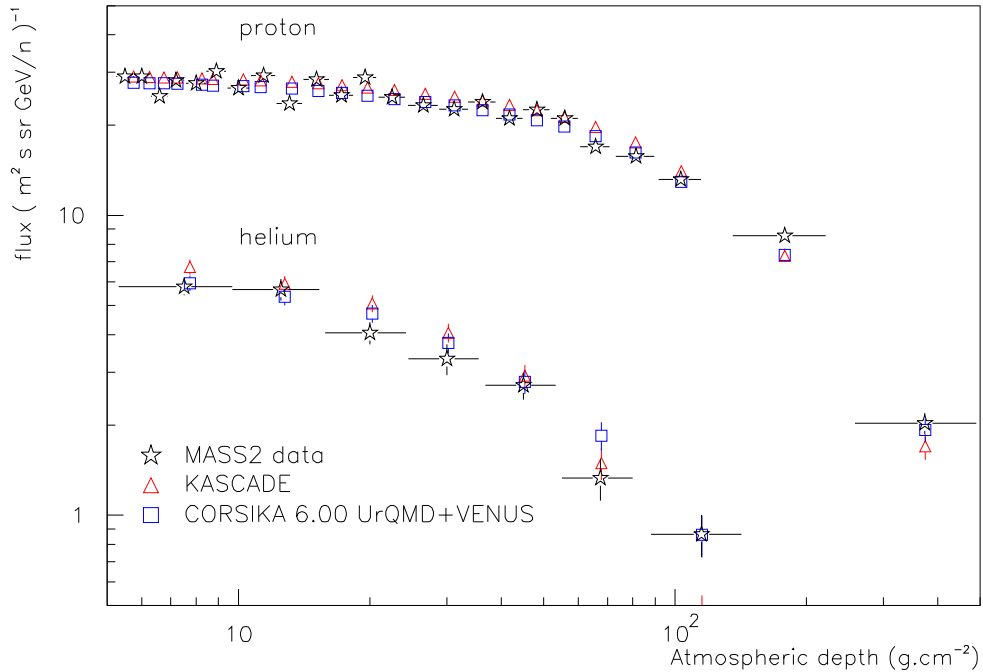}
\caption{ 
Helium and proton fluxes as a function of atmospheric depth measured 
by MASS2  \cite{mass2protons} (stars) compared with simulations. 
Proton kinetic energy is between 4.1 and 14.1 GeV, and helium between 1.7 and 9.1 GeV/n.
}\label{primarydepth}  
\end{figure}

All models are in good agreement with the variation of the proton flux as a function of 
atmospheric depth as measured by MASS2.
The best agreement is obtained with KASCADE and CORSIKA 6.00 with UrQMD, shown in figure 
\ref{primarydepth}. For the helium flux, the results are also satisfactory except for CORSIKA version 
5.60 running GHEISHA. With this model, we observe a quantitative discrepancy with experimental data 
as the helium flux hardly decreases with atmospheric depth. This gives rise to huge $\chi^{2}$ values
in table \ref{table_chi2} for this model. This effect is corrected in version 6.00.
 KASCADE and CORSIKA 6.00 with UrQMD helium fluxes are also shown figure  \ref{primarydepth}.

\subsection{Muons}
 
\begin{figure}
\includegraphics[width=14.cm]{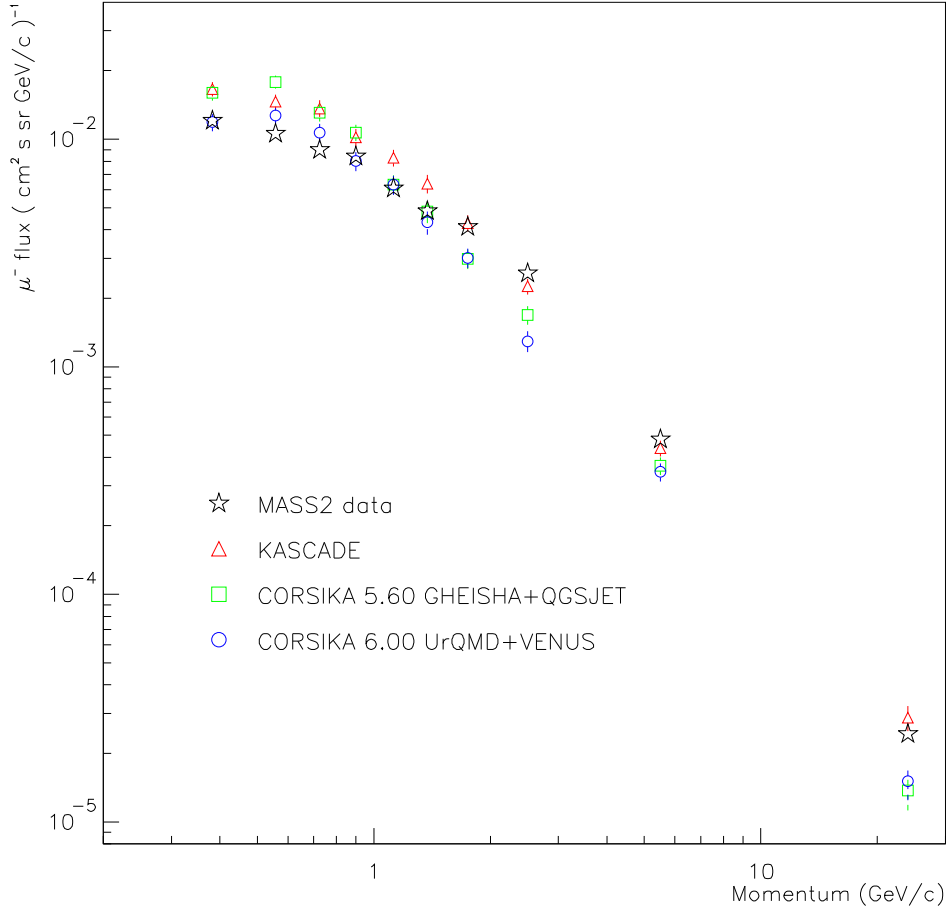}
\caption{$\mu^{-}$ momentum spectra measured by MASS2 \cite{mass2muons} 
(stars) in the atmospheric depth interval 164 and 255 g.cm$^{-2}$ 
compared with simulations.
}\label{muonspectrum} 
\end{figure}

The $\mu^-$ spectra for different atmospheric depth ranges are estimated. 
Figure \ref{muonspectrum} shows the MASS2 observed spectrum 
in the atmospheric depth between 164 and 255 g.cm$^{-2}$ along with the 
prediction of some models. The slope obtained with CORSIKA with GHEISHA is slighty 
steeper than the experimental one. 
KASCADE and CORSIKA 6.00 with UrQMD fit well the observed spectra.

When comparing the muon flux as a function of atmospheric depth for different energy ranges, the discrepancies appear more clearly.
In figure \ref{muondepth} one can see that GHEISHA in CORSIKA 5.60 does not give enough muons at high momenta 
(above 4 GeV/c). On the same figure, we see that KASCADE and CORSIKA 6.00 with UrQMD are in good agreement.

\subsection{Summary}

It turns out that KASCADE hadronic model gives good estimations of the proton, helium 
and muon fluxes and spectra at various atmospheric depths. 
Concerning CORSIKA, results vary from one hadronic model to another. Though in this 
analysis we do not intend to test the high-energy interaction 
models, we can give some conclu\-sions concerning the low-energy ones. 
ISOBAR is ruled out (it is not included in version 6.00),
 GHEISHA gives rise to too many low-energy muons (steeper spectrum) especially in 
version 6.00, and the model which best fits the data is UrQMD. 
    
%%%%%%%%%%%%%%%%%%%%%%%%%%%%%%%%%%%%%%%%%%%%%%%%%%%%%%%%%%%%%%%%%%%%%%%%%%%%%%%%%%%%%%%%%%%%%%%%%%%%%%%%%%%%

\begin{table*}[hbt]
\begin{tabular*}{\textwidth}{@{}l@{\extracolsep{\fill}}lllll}
\hline 
Generator & \multicolumn{2}{c}{Hadronic model}  & proton        & helium       & muon     \\
          & low energy     & high energy  & $\chi^2 / 24$ & $\chi^2 / 7$ & $\chi^2 / 70$  \\\hline

KASCADE  & \multicolumn{2}{c}{Gaisser and Stanev} & 1.6 $\dag$ & 1.3 $\dag$ & 2.5 $\ddag$ \\ %ok
  
CORSIKA 5.60 & ISOBAR & HDPM    &	5.0 & 1.6 & 16.5\\ %ok
CORSIKA 5.60 & GHEISHA & HDPM   &	2.8 & 63  & 3.8	\\ %ok
CORSIKA 5.60 & GHEISHA & QGSJET &	2.8 & 53 & 3.4 $\ddag$ \\ %ok
CORSIKA 5.60 & GHEISHA & VENUS  &	1.8 & 56  & 4.2	\\ %ok
CORSIKA 5.60 & GHEISHA & SIBYLL &	4.6 & 59  & 3.5 \\ %ok
CORSIKA 5.60 & GHEISHA & DPMJET & 2.1 & 67  & 2.7 \\ %ok

CORSIKA 6.00 & GHEISHA & HDPM   &	1.7 & 5.5 & 9.6	\\ %ok
CORSIKA 6.00 & GHEISHA & QGSJET &	3.2 & 2.7 & 8.0	\\ %ok
CORSIKA 6.00 & GHEISHA & VENUS  &	2.0 & 4.7 & 10.1\\ %ok
CORSIKA 6.00 & GHEISHA & SIBYLL &	1.9 & 3.4 & 9.9	\\ %ok
CORSIKA 6.00 & GHEISHA & DPMJET &	2.3 & 8.4 & 9.5 \\ %ok
CORSIKA 6.00 & GHEISHA & NEXUS  &	2.4 & 5.9 & 9.3 \\ %ok

CORSIKA 6.00 & URQMD & HDPM     &	1.8 & 5.4 & 3.5	\\ %ok
CORSIKA 6.00 & URQMD & QGSJET   &	1.8 & 5.2 & 2.0	\\ %ok
CORSIKA 6.00 & URQMD & VENUS    &	1.3 $\dag$ & 4.4 $\dag$ & 2.5 $\ddag$ \\ %ok
CORSIKA 6.00 & URQMD & SIBYLL   &	1.5 & 3.8 & 2.4	\\ %ok
CORSIKA 6.00 & URQMD & DPMJET   &	1.5 & 3.1 & 2.5	\\ %ok 
CORSIKA 6.00 & URQMD & NEXUS    &	1.8 & 2.5 & 1.7 \\ %ok 
  
\hline
\end{tabular*}
{$\dag$ {\em see Fig. \ref{primarydepth}}}, {$\ddag$ {\em see Fig. \ref{muondepth}}}
\caption{
Global $\chi^2$ results per degree of freedom from comparison of hadronic models with MASS2 
data  \cite{mass2protons}  \cite{mass2muons} concerning proton, helium (figure \ref{primarydepth}) 
and $\mu$ (figure \ref{muondepth}) fluxes as a function of atmospheric depth.
}\label{table_chi2} 
\end{table*}

%%%%%%%%%%%%%%%%%%%%%%%%%%%%%%%%%%%%%%%%%%%%%%%%%%%%%%%%%%%%%%%%%%%%%%%%%%%%%%%%%%%%%%%%%%%%%%%%%%%%%%%%%%%%

\section{Uncertainties}

%%%%%%%%%%%%%%%%%%%%%%%%%%%%%%%%%%%%%%%%%%%%%%%%%%%%%%%%%%%%%%%%%%%%%%%%%%%%%%%%%%%%%%%%%%%%%%%%%%%%%%%%%%%%

The first uncertainty of the simulation is  the cosmic rays spectrum slope which is about $-2.7\pm0.06$ for nuclei. 
We have performed the same study with the energy spectrum slope varying within the range of its uncertainties.
The differences obtained are found to be smaller than the statistical errors of the simulation. 
They are about $3\%$ for protons and 
 $10\%$ for helium at about 50 g.cm$^{-2}$, and  $5\%$ for $p_{\mu}< 0.9$ GeV/c and $10\%$ for 
$p_{\mu}> 4$ GeV/c at about 100 g.cm$^{-2}$ for muons.
Hence, since MASS2 errors are of the same order, an upper limit of the variation of the $\chi^2$ 
value per degree of freedom with the primary spectrum slope
 is 0.5 which does not modify the conclusions concerning the low energy hadronic models.
  
Another uncertainty comes from the normalization of the muon flux with respect to the proton flux in MASS2. 
It is due to the event selection 
criteria in their analysis which are not the same for protons and muons. This was checked \cite{mass2muons} 
and this uncertainty on the ratio of selection
efficiency between proton and muons does not exceed $2\%$.

In all the simulations performed in this study, an homogeneous magnetic field is assumed between the injection altitude 
and the ground whereas the actual geomagnetic field decreases with altitude within a 2.5\% range. 
This approximation largely dominates the uncertainty on the magnetic field value on ground used in the models. 

Another source of error may come for our atmospheric depth profile parametrization. On the one hand the 
data from the US Radiosonde database \cite{radiosonde} are a mixture of measurements and models, and there 
is no data available for the Fort Sumner location and on the other hand the simulation codes require a 
parametrization of these data (several layers with $\rho(h)=a_i \times e^{-h/b_i}$). In order to quantify 
the impact of these uncertainties, we have 
performed the same simulation with two sets of atmospheric profile parametrization (fitted from Midland-Odessa 
and Albequerque data). The discrepancy between the global $\chi^2$ obtained is about $1\%$ but it does 
not affect the conclusions of this analysis.

%%%%%%%%%%%%%%%%%%%%%%%%%%%%%%%%%%%%%%%%%%%%%%%%%%%%%%%%%%%%%%%%%%%%%%%%%%%%%%%%%%%%%%%%%%%%%%%%%%%%%%%%%%%%

\section{Conclusion}

%%%%%%%%%%%%%%%%%%%%%%%%%%%%%%%%%%%%%%%%%%%%%%%%%%%%%%%%%%%%%%%%%%%%%%%%%%%%%%%%%%%%%%%%%%%%%%%%%%%%%%%%%%%%
Thanks to MASS2 balloon flight cosmic-ray measurements, a constrained test of the low-energy hadronic 
models of air-shower simulations is possible. 

We have compared KASCADE and CORSIKA simulations with data. 
KASCADE gives rise to a proton, helium and muon flux evolution over atmospheric depth in agreement with 
experimental data. 
In CORSIKA, GHEISHA gives a steeper muon spectrum whereas the agreement is satisfactory with UrQMD.
However we would like to stress the fact that this analysis only tests the mean longitudinal development 
of air-showers and for instance 
does not provide information on the lateral spread of the shower.

The conclusions of this analysis are confirmed by the new results from CAPRICE98, a similar balloon 
experiment launched during spring 1998 on the same site \cite{caprice98}. 
The results from BESS \cite{bess1,bess2} could also be used to continue this study for example to test
another atmospheric condition.
    
%%%%%%%%%%%%%%%%%%%%%%%%%%%%%%%%%%%%%%%%%%%%%%%%%%%%%%%%%%%%%%%%%%%%%%%%%%%%%%%%%%%%%%%%%%%%%%%%%%%%%%%%%%%%

\begin{ack}

%%%%%%%%%%%%%%%%%%%%%%%%%%%%%%%%%%%%%%%%%%%%%%%%%%%%%%%%%%%%%%%%%%%%%%%%%%%%%%%%%%%%%%%%%%%%%%%%%%%%%%%%%%%%
We wish to thank John M. Hobbie from the National Scientific Balloon Facility for providing us the atmosphere 
profiles in New Mexico, J. Wentz and K. Bernl\"ohr for their advice concerning CORSIKA, M. Circella for his 
references relating to cosmic rays balloon measurements and Michael Punch for his comments.  
%%%%%%%%%%%%%%%%%%%%%%%%%%%%%%%%%%%%%%%%%%%%%%%%%%%%%%%%%%%%%%%%%%%%%%%%%%%%%%%%%%%%%%%%%%%%%%%%%%%%%%%%%%%%

\end{ack}

\end{document}